


\magnification\magstep1
\parskip=\medskipamount
\hsize=6 truein
\vsize=8.0 truein
\hoffset=.2 truein
\voffset=0.4truein
\baselineskip=14pt


\font\titlefont=cmbx12
 at 10 truept
\font\authorfont=cmcsc10
\font\addressfont=cmsl10 at 10 truept
\font\smallbf=cmbx10 at 10 truept
 4



\outer\def\beginsection#1\par{\vskip0pt plus.2\vsize\penalty-150
\vskip0pt plus-.2\vsize\vskip1.2truecm\vskip\parskip
\message{#1}\leftline{\bf#1}\nobreak\smallskip\noindent}


\outer\def\subsection#1\par{\vskip0pt plus.2\vsize\penalty-80
\vskip0pt plus-.2\vsize\vskip0.8truecm\vskip\parskip
\message{#1}\leftline{\it#1}\nobreak\smallskip\noindent}


\newcount\notenumber

\def\note{\advance\notenumber by 1
\footnote{$^{\the \notenumber}$}}      


\newdimen\itemindent \itemindent=13pt
\def\textindent#1{\parindent=\itemindent\let\par=\resetpar%
\indent\llap{#1\enspace}\ignorespaces}

\let\oldpar=\par
\def\resetpar{\oldpar\parindent=0pt\let\par=\oldpar}

\font\ninerm=cmr9 \font\ninesy=cmsy9
\font\eightrm=cmr8 \font\sixrm=cmr6
\font\eighti=cmmi8 \font\sixi=cmmi6
\font\eightsy=cmsy8 \font\sixsy=cmsy6
\font\eightbf=cmbx8 \font\sixbf=cmbx6
\font\eightit=cmti8
\def\eightpoint{\def\rm{\fam0\eightrm}
  \textfont0=\eightrm \scriptfont0=\sixrm \scriptscriptfont0=\fiverm
  \textfont1=\eighti  \scriptfont1=\sixi  \scriptscriptfont1=\fivei
  \textfont2=\eightsy \scriptfont2=\sixsy \scriptscriptfont2=\fivesy
  \textfont3=\tenex   \scriptfont3=\tenex \scriptscriptfont3=\tenex
  \textfont\itfam=\eightit  \def\it{\fam\itfam\eightit}%
  \textfont\bffam=\eightbf  \scriptfont\bffam=\sixbf
  \scriptscriptfont\bffam=\fivebf  \def\bf{\fam\bffam\eightbf}%
  \normalbaselineskip=9pt
  \setbox\strutbox=\hbox{\vrule height7pt depth2pt width0pt}%
  \let\big=\eightbig \normalbaselines\rm}
\catcode`@=11 %
\def\eightbig#1{{\hbox{$\textfont0=\ninerm\textfont2=\ninesy
  \left#1\vbox to6.5pt{}\right.\n@space$}}}
\def\vfootnote#1{\insert\footins\bgroup\eightpoint
  \interlinepenalty=\interfootnotelinepenalty
  \splittopskip=\ht\strutbox %
  \splitmaxdepth=\dp\strutbox %
  \leftskip=0pt \rightskip=0pt \spaceskip=0pt \xspaceskip=0pt
  \textindent{#1}\footstrut\futurelet\next\fo@t}
\catcode`@=12 %




\def\shalf{\hbox{${\textstyle{1\over 2}}$}}
\def\quarter{\hbox{${\textstyle{1\over 4}}$}}
\def\threequarter{\hbox{${\textstyle{3\over 4}}$}}

\def\twothird{\hbox{${\textstyle{2\over 3}}$}}
\def\eightthird{\hbox{${\textstyle{8\over 3}}$}}
\def\l{\tau}

\tolerance=500


\rightline{Freiburg, THEP-95/13}
\rightline{gr-qc/9508040}
\bigskip
{\baselineskip=24 truept
\titlefont
\centerline{A EUCLIDEAN BIANCHI MODEL BASED ON $S^3/D_8^*$}
}

\vskip 1.1 truecm plus .3 truecm minus .2 truecm

\centerline{\authorfont Domenico Giulini\footnote*{
e-mail: giulini@sun2.ruf.uni-freiburg.de}}
\vskip 2 truemm
{\baselineskip=12truept
\addressfont
\centerline{Fakult"t f\"ur Physik,
Universit\"at Freiburg}
\centerline{Hermann-Herder Strasse 3, D-79104 Freiburg, Germany}
}
\vskip 1.5 truecm plus .3 truecm minus .2 truecm

\centerline{\smallbf Abstract}
\vskip 1 truemm
{\baselineskip=12truept
\leftskip=3truepc
\rightskip=3truepc
\parindent=0pt

{\eightpoint
We explain how the round four sphere can be sliced along homogeneous
3~-~manifolds of topology $S^3/D_8^*$. This defines a Euclidean Bianchi
type IX model for Einstein's equations with cosmological constant.
The geometric properties of this model are investigated.
\par}}

\noindent
PACS: 02.40.-k, 02.40.Ma, 98.80.Hw\hfill\break
 MSC:  53C42

\beginsection Introduction

Spacetimes with homogeneous spatial slices are used extensively
as testing ground for Einstein's field equations of
general relativity [1-2]. Technically speaking, spatial homogeneity is
used to truncate all the infinitely many inhomogeneous degrees
of freedom thus keeping only the finitely many homogeneous ones.
Analytically the field equations then pose a much simpler problem
that can be approached by methods familiar from classical or
quantum mechanics. Physically speaking,
such a drastic procedure seems justified provided one confines
interest to cosmological scales. Classically a spacetime must
clearly be a Lorentzian manifold. However, some
attempts to quantize gravity make essential use of Euclidean
spacetimes, that is, spacetimes with a Riemannian metric. We refer
to [3-5] for more details and motivation.
The vacuum Einstein equations then simply require the Euclidean
spacetime to be an Einstein manifold. For identical reasons as
in the Lorentzian case, one is still interested in a homogeneous
slicing. Although the specification `spatial' does not make any
sense in the Euclidean context, it is sometimes still used as a
reminder on its use for the Lorentzian case\note{For example,
in quantum cosmology one considers manifolds which carry a
Riemannian metric on one side of a spatial slice and a Lorentzian
metric on the other [5].}. In this paper
we consider the Einstein manifold $S^4$ with its standard round
metric. But instead of slicing it along three dimensional spheres
[4] we will slice it along homogeneous manifolds
$S^3/D_8^*$. Here $D_8^*$ is the binary dihedral group of eight
elements. As a subgroup of $SU(2)$ it is easily expressed using
the unit quaternions $1,i,j,k$, $i^2=j^2=k^2=-1$, $ij=k$ and cyclic.
Then: $D_8^*=\{\pm 1,\pm i,\pm j,\pm k\}$. In other words, we
consider a Bianchi type IX model [2] with
slightly more complicated topology than usual.

We investigate in a self
contained fashion the geometry of these isoparametric embeddings,
which is indeed quite interesting. The plan of the paper is as
follows: In section 1 we show in an explicit and quite elementary
fashion how to construct an isometric family of embeddings
$S^3/D_8^*\hookrightarrow S^4$. In section 2 we determine the
corresponding family of 3-dimensional metrics and discuss their
geometry. We end with a brief discussion. An appendix provides
some background material on the geometry of left invariant metrics
on $S^3/G$, where $G$ can be any finite subgroup of $SU(2)$. Such
geometries are of Bianchi class IX.

\beginsection{Section 1: Isoparametric embeddings of $S^3/D^*_8$
                         into $S^4$}

A one-parameter family of embeddings is called isoparametric, if there is an
embedding parameter $t$ that labels the different leaves, such that the
extrinsic geometry is homogeneous along a leaf and thus depends only on $t$.
Quantities like the principal curvatures or the norm of $dt$ are just
functions of $t$. This section deals with the construction of a particular
such family of embeddings of a certain 3-manifold into the round unit four
sphere.

Let $A$ be an element of $SM_3$, the space of real symmetric $3\times 3$
matrices. If identified with $R^6$ via
$$
A=\pmatrix{x_1&{1\over\sqrt 2}x_2&{1\over\sqrt 2}x_3\cr {1\over\sqrt 2}x_2&
x_4&{1\over\sqrt 2}x_5\cr {1\over\sqrt 2}x_3&{1\over\sqrt 2}x_5&x_6\cr}
$$
we see that the two conditions
$$ \eqalignno{
1.)&\quad tr(AA)=1\qquad R^6\rightarrow S^5   &(1.1)\cr
2.)&\quad tr(A)=0\qquad\quad S^5\rightarrow S^4    &(1.2)\cr}
$$
reduce the manifold of corresponding matrices to the unit five-, respectively
unit four sphere. On $R^6$ we have the Euclidean metric
$$
ds^2= tr(dA\cdot dA)
\eqno(1.3)
$$
where the dot ``$\cdot$'' indicates matrix- and symmetrized tensor product.
It induces the standard metric of the unit four-sphere given by the
conditions (1.1) and (1.2).

On $SM_3$ there is an action of $SO(3)$ by conjugation
$$\eqalignno{
SO(3)\times SM_3&\rightarrow SM_3   \cr
(R,A)&\mapsto RAR^{-1}   &(1.4)            \cr}
$$
which by (1.1) (1.2) and (1.3) is seen to induce an isometry on $S^4$.
We are interested in the orbits of this action. Using the fact that the
exponential map provides a homeomorphism between the space of symmetric
matrices, $SM_3$, and the space of positive definite symmetric matrices,
$PSM_3$:
$$\eqalign{
SM_3&\mathop{\longrightarrow}^{\rm homeo}PSM_3\cr
A&\mathop{\longmapsto}\exp(A)\cr}
$$
we may think of the eigenvalues as the diameters of an
ellipsoid\note{The whole purpose of
using the homeomorphic space $PSM_3$ is to introduce non-negative diameters
and therefore achieve visualization. What really matters is whether the
diagonal elements are pairwise distinct or whether two coincide.}.
The group $SO(3)$ then still acts on $PSM_3$ by conjugation leaving invariant
the image of $S^4$ in $PSM_3$ under $\exp$. That these matrices are
$SO(3)$-diagonalisable means that we can chose a suitable point on each orbit
such that $A$ and hence $\exp (A)$ is diagonal. There are six such points on
each generic orbit, due to the permutations of the principal axes generated by
$\pi/2$-rotations about any two different such axes.
In order to fix this redundancy, we assume the eigenvalues to be ordered. For
later convenience we choose
$$
\lambda_2\geq\lambda_1\geq\lambda_3
\eqno{(1.5)}
$$
If we continue to refer to the general point
in $SM_3$ as $A$ and the diagonal point as $\Lambda$, we have
$$\eqalignno{
\Lambda&=\hbox{diag}\{\lambda_1,\lambda_2,\lambda_3\}&(1.6)  \cr
\exp(\Lambda)&=\hbox{diag}\{e^{\lambda_1},e^{\lambda_2},e^{\lambda_3}\}&(1.7)
\cr}
$$
We visualize the space of matrices $PSM_3$ as the configuration space of
the corresponding ellipsoid with diameters
along the principal axes of $e^{\lambda_1},e^{\lambda_2},e^{\lambda_3}$.
Since the $SO(3)$
action is just the standard rotation of this ellipsoid, we can immediately
read off the stabilizer groups. We have two cases corresponding to the
principal diameters being pairwise distinct (case 1), or whether two of them,
say the first and the second, coincide (case 2). The case where all three
coincide is excluded by (1.1) and (1.2).

Case 1: The stabilizer group is generated by $\pi$-rotations about any
two of the principal axes. The group thus generated
is\note{$D_{2n}$ is the subgroup of $SO(3)$ corresponding to the symmetries
of the n-prism, $D^*_{4n}$ is its 2-fold cover group in $SU(2)$. The subscript
denotes the order of the group.}
$$
Z_2\times Z_2\cong D_4=\{\pm1,\pm i,\pm j,\pm k\}/Z_2=D^*_8/Z_2
\eqno{(1.8)}
$$
where $1,i,j,k$ denote the unit quaternions. Note that $D^*_8$ is the
pre-image under the identification homomorphism (antipodal map) ${\cal A}$
$$
S^3\cong SU(2)\mathop{\longrightarrow}^{\cal A} SO(3)\cong RP^3
\eqno{(1.9)}
$$
The orbit type is therefore given by
$$
SO(3)/D_4=SU(2)/D^*_8\cong S^3/D^*_8
\eqno{(1.10)}
$$
This manifold can be visualized by taking a solid cube whose opposite sides
are pairwise identified after a relative $\pi/2$ - rotation about
the axis through their midpoints. It is obvious from the construction that
its fundamental group is $D^*_8$, the higher ones being just those of its
covering $S^3$. Its homology (resp. cohomology) is
$H_*=(Z,Z_2\times Z_2,0,Z)$ (resp. $H^*=(Z,0,Z_2\times Z_2,Z)$).

Case 2: Here the third principal axis is a symmetry axis. The stabilizer
group is generated by all rotations about the symmetry axis and a $\pi$-
rotation about any other axis perpendicular to it. The orbit manifold can
therefore be identified with the configuration space
of the symmetry axis, i.e. by the space of all lines through the origin, in
other words $RP^2$. A more formal way to see this is to
embed the acting SO(3) in $O(3)$\note{The point of doing so is that, in
contrast to $SO(3)$, the stabilizer group is now the product of subgroups of
$O(3)$.}. The stabilizer is then $O(2)\times Z_2$ where the $Z_2$ is generated
by reflections at the origin. $O(2)$ contains reflections about the xz- and
yz-plane. Since $Z_2$ is normal it acts on the quotient $O(3)/O(2)\cong S^2$,
just by identifying opposite points. Again we get $RP^2$.

Each of the orbits is an isometrically embedded submanifold of $S^4$.
So we found isometric embeddings of $S^3$ resp. $RP^2$ into $S^4$. The latter
one turns out to be congruent to the celebrated Veronese embedding (section 2).
If composed with the stereographic
projection based at a point in the complement of the orbit in question, we
obtain a conformal embedding of $S^3/D^*_8$ resp. $RP^2$ into $R^4$. We will
comment on this at the end of section 2.

By (1.1) and (1.2), the $\lambda_i$'s just parameterize a great circle of
intersection
of the 2-sphere with a plane whose normal is $n={1\over\sqrt{3}}(1,1,1)$.
This normal is obtained from $e_3$ by applying the rotation
$K=R_{z}(\pi/4)\circ R_{y}(\cos^{-1}({1\over \sqrt{3}}))$ i.e.
$$
K=\pmatrix{{1\over\sqrt{6}}&-{1\over\sqrt{2}}&{1\over\sqrt{3}}\cr
            {1\over\sqrt{6}}&{1\over\sqrt{2}}&{1\over\sqrt{3}}\cr
           -\sqrt{2\over 3}&0&{1\over\sqrt 3}\cr}
\eqno{(1.11)}
$$
In the rotated frame, $e'_i=K^j_{{} i}e_j$, the great circle has the coordinate
representation $\vec x(t)=(\cos t,\sin t,0)$ so that we find
$$
\eqalignno{
\lambda_1(t)&=\sqrt{2\over 3}\cos (t+{\pi\over 3})\cr
\lambda_2(t)&=\sqrt{2\over 3}\cos(t-{\pi\over 3})&(1.12) \cr
\lambda_3(t)&=\sqrt{2\over 3}\cos(t+\pi).  \cr}
$$
For later application we also note:
$$
\eqalignno{
\lambda_1-\lambda_2&=-\sqrt{2}\sin t\cr
\lambda_2-\lambda_3&=-\sqrt{2}\sin(t-{2\over 3}\pi)  &(1.13)\cr
\lambda_3-\lambda_1&=-\sqrt{2}\sin(t+{2\over 3}\pi)\,.\cr}
$$
Condition (1.5) then restricts the range of $t$ to $t\in [0,\pi/3]$ where
interior points correspond to generic orbits of topology $S^3/D^*_8$ and
boundary points ($t=0,t=\pi/3$) to degenerate orbits of topology $RP^2$.
In the sequel we set $S^3/D^*_8=:
\Sigma$. We have thus found a one parameter family of embeddings
$$
\eqalignno{
\phi_t:\, \Sigma &\longrightarrow S^4  \qquad t\ \in\,]0,\pi/3[ &(1.14) \cr
          \Sigma &\longmapsto \Sigma_t\subset S^4  \cr}
$$
where $\Sigma_t$ denotes the image of $\Sigma$ under $\phi_t$ in $S^4$.

$\Sigma$ is only a special example of a so called {\it space form},
which are defined to be the quotient spaces of $S^3$ with respect to a freely
acting finite subgroup $G$ of $SO(4)$. Its action on $S^3$ is best understood
using
$$
SO(4)\cong {SU(2)\times SU(2)\over Z_2}
\eqno{(1.15)}
$$
where the $Z_2$ is generated by $(-1,-1)$. Elements in $SO(4)$ are then
written as $Z_2$-equivalence classes $[g,h]$ with $g,h\in SU(2)$.
If we then identify $S^3$ with $SU(2)$, the action is given by:
$$
\eqalignno{
S^3\times SO(4)&\longrightarrow  S^3 \cr
\left\{p,[g,h]\right\}&\longmapsto g\cdot p\cdot h^{-1} &(1.16)\cr}
$$
Equation (1.15) suggests to call $[\pm 1,SU(2)]=:SU(2)_R$ the right $SU(2)$
and $[SU(2),\pm 1]=:SU(2)_L$ the left $SU(2)$. Clearly $SU(2)_R\cap SU(2)_L
=\left\{[1,1],[1,-1]\right\}$. If $G$, besides being a finite subgroup of
$SO(4)$, is also a non-cyclic (i.e. not a $Z_p$) finite subgroup of $SU(2)$
it must sit in either $SU(2)_R$ or $SU(2)_L$ in order to act freely on $S^3$.
In these cases the resulting quotient space is a {\it homogeneous
space form} and $G$ can w.l.o.g. be taken as a subgroup
of
$SU(2)_R$\note{If we call $\Sigma_R$ (resp. $\Sigma_L$) the quotient manifold
with respect to the right
(resp. left) $SU(2)$ action with points $[p]_R$ (resp. $[p]_L$),
$p\in S^3$, then they are diffeomorphically related by
$[p]_R\mapsto [p^{-1}]_L$}.
If we started with the standard round metric on
$S^3$ (the bi-invariant metric on $SU(2)$), the residual isometry group of
$S^3/G$ would be
$$
\hbox{Isom}(S^3/G)=SO(3)\times N_{G}(SU(2))/G
\eqno{(1.17)}
$$
where $N_G(SU(2))$ denotes the normalizer of $G$ in $SU(2)$. All these
spherical
space forms can be obtained by suitably identifying opposite sides of some
polyhedron (the fundamental domain). It is not hard to show that the symmetry
group of this polyhedron is just the normalizer of $G/Z_2$ (the image of $G$
under the quotient map $\cal A$ of eq. (1.9)) in the diagonal $SO(3)$ of
$SO(4)$. In the case at hand, $G=D^*_8$ and we already noted that the
fundamental domain for $\Sigma$ is a cube. We have $N_{D^*_8}(SU(2))=O^*$ and
hence
$$
\hbox{Isom}(\Sigma)=SO(3)\times P_3
\eqno{(1.18)}
$$
where $P_3$ is the six element permutation group of three objects. The
diagonal $P_3$ in $\hbox{Isom}(\Sigma)$ can then be identified with the
permutations of the principle axes of the ellipsoid in the previous
discussion.

Clearly, we can not expect the orbits $\Sigma_t$ to have this maximally
symmetric metric, albeit we know by construction that the isometry group
must contain a left acting SO(3). There is however no reason for the
$\pi/2$-rotations about principal axes to be isometries as well. As a minimal
symmetry we therefore expect the generic orbits to have a metric that, if
lifted to $S^3$ ($\cong SU(2)$), is left invariant and whose right isometries
are equal to $D^*_8$.
The next section contains an explicit calculation of the orbit metric
showing that the metric is indeed of the minimal symmetric form.

Finally we make some topological remarks. Since the unit quaternion $k$ (say)
generates a normal $Z_4$-subgroup of $D^*_8$, we can take the quotient
$S^3\rightarrow \Sigma$ stepwise; first with respect to the $Z_4$, then
dividing the remaining quotient $Z_2$. After the first step we arrive at the
lens space $L(4,1)$ with its well known structure as a principal $U(1)$-bundle
over $S^2$. The next step transforms this into a circle bundle over $RP^2$,
which is clearly not a principal bundle since the (right) $U(1)$ which
acted on $L(4,1)$ does not act on the $Z_2$-quotient of it.
$\Sigma$ is thus a circle bundle over $RP^2$, and if we cut $S^4$ along a
$\Sigma_t$ each piece is a 2-disc bundle over $RP^2$ where the zero section
can be identified with Veronese's $RP^2$ in $S^4$.

\beginsection{Section 2. The metric}

Let us compute the metric by coordinatizing the orbit direction by the group
parameters and the perpendicular direction by the single parameter that
labels the range of the eigenvalues $\lambda_1,\lambda_2, \lambda_3$
obeying (1.1), (1.2) and (1.5). We thus write
$$
A=R\Lambda R^{-1}
\eqno{(2.1)}
$$
This implies that our $SO(3)$-action is given by left translations
on the $R$-factor, whereas rotations of the principal axes (by which the
identification $SO(3)\rightarrow \Sigma$ is made) correspond to right
translations. A short calculation then leads to
$$
ds^2=tr(dA\cdot dA)=tr\left\{d\Lambda \cdot d\Lambda
+2\left[(R^{-1}dR\Lambda)^2 - (R^{-1}dR)^2\Lambda^2 \right]\right\}
\eqno{(2.2)}
$$
Introducing the basis\note{Whereas right invariant sections over $SO(3)$
induce well defined sections on $\Sigma$, this holds in the left invariant
case only for even-valence tensors, like e.g. the metric (2.8). The odd-valence
sections are well defined only locally, and globally up to sign. In the sequel
we still refer to the basis of left invariant 1-forms, or their dual basis of
left invariant vector fields without explicit mention of this sign ambiguity.
Note that the notion of integral lines of left invariant vector fields still
makes unambiguous sense.} of left invariant 1-forms via
$$
R^{-1}dR=:T^al_a\,,
\eqno{(2.3)}
$$
where the basis of $so(3)$, the Lie algebra of $SO(3)$, is given by
$$
(T^a)^b_{\, c}=\varepsilon_{abc}\,,\quad\hbox{s.t.}\quad \left[T^a,T^b\right]
=-\varepsilon_{abc}T^c
\eqno{(2.4)}
$$
one explicitly finds, using Euler angles $(\psi,\theta,\varphi)$,
$$
\eqalignno{
l_1&=\cos\psi\sin\theta\,d\varphi-\sin\psi\,d\theta &(2.5.a) \cr
l_2&=\sin\psi\sin\theta\,d\varphi+\cos\psi\,d\theta &(2.5.b) \cr
l_3&=d\psi+\cos\theta\,d\varphi                     &(2.5.c) \cr
\hbox{obeying}\quad dl_a&=\shalf
\varepsilon_{abc}l_b\wedge l_c\,.                   &(2.6)\cr}
$$
Inserting (1.6) and (2.3) into (2.2) gives:
$$
ds^2=(d\lambda_1^2+d\lambda_2^2+d\lambda_3^2)
    +2\left[(\lambda_2-\lambda_3)^2\, l_1^2+(\lambda_3-\lambda_1)^2 \, l_2^2
     + (\lambda_1-\lambda_2)^2\, l_3^2\right]
\eqno{(2.7)}
$$
Using (1.12) and (1.13) we get the final form of the metric:
$$
ds^2=dt^2+4\left[\sin^2 (t-\twothird\pi)\ l_1^2+
                           \sin^2 (t+\twothird\pi)\ l_2^2+
                           \sin^2 (t)\ l_3^2 \right]
\eqno{(2.8)}
$$
For small values of $t$ this expression becomes:
$$
ds^2=dt^2+4t^2\,l_3^2+3(d\theta^2+\sin^2\theta\,d\varphi^2)
$$
The regularity at $t=0$, though obvious from the whole geometry, is now
explicitly seen due
to the identification via the right $\exp(T_3\pi)$ translation\note{
Right translations are generated by left invariant vector fields. The left
invariant vector field corresponding to $T_3$ is ${\partial\over\partial\psi}$
hence $\psi$ has range $0\leq\psi\leq\pi$}. More importantly, the degenerated
orbit, $RP^2$, is seen to carry a metric of constant sectional curvature $1/3$
and must therefore be congruent to the Veronese embedding. Clearly, all this
holds equally well for $t={\pi\over 3}$.

Let us now focus on the generic orbits. On the universal covering, $S^3$, the
Euler angles have the range $0\leq\psi\leq 4\pi$, $0\leq
\theta\leq \pi$ and $0\leq \varphi\leq 2\pi$. A typical orbit will be covered
eight times. The volume of the typical orbit then follows:
$$\eqalignno{
V&\,=\,\vert\sin (t)\,\sin (t-\twothird\pi)\,
\sin (t+\twothird\pi)\vert\,\int_{S^3} l_1\wedge l_2\wedge l_3  \cr
&\,=\,\vert\sin (t)\,\sin (t-\twothird\pi)\,\sin (t+\twothird\pi)
\vert\quad 8\ (2\pi^2) \cr
&\,=\,\sin (t)\, \left[\cos^2 (t)-\quarter\right ]
\,8\ (2\pi^2)\,=\,\sin (t)\,\left[\threequarter
-\sin^2 (t)\right]\, 8\  (2\pi^2) &(2.9)\cr
&\,=\,\vert\lambda_1-\lambda_2\vert
\vert\lambda_2-\lambda_3\vert \vert \lambda_3 -\lambda_1\vert\, 2\sqrt{2}\,
(2\pi^2) &(2.10) \cr}
$$
The maximal value is reached for $t={\pi\over 6}$ and given by
$$
V^{(3)}_{\hbox{max}}= 4\pi^2
\eqno{(2.11)}
$$
which is twice the volume of the equatorial 3-sphere in the unit $S^4$.
As a check, we calculate the total 4-dimensional volume from (2.8) and find
$$\eqalign{
V^{(4)}&=16\pi^2\int_0^{\pi\over 3}dt\ \sin (t)
\left[\cos^2 t-\quarter\right]
\cr
&=16\pi^2\int_{1\over 2}^1 dz\ \left[z^2-\quarter\right]=
\eightthird\pi^2 \cr}
$$
which is just the volume of the unit 4-sphere as it must be. Although (2.11)
is just the maximal volume amongst the particular 1-parameter variation
(2.9), general
arguments\note{The argument runs as follows: Let $K$ be the
trace of the extrinsic curvature tensor, and $\nu$ the normal vector field
to the embeddings $\phi_t$. Fix $t_1$ such that $\phi_{t_1}(\Sigma)
=\Sigma_{t_1}$ is a generic orbit. Define a section $\eta$ in the normal bundle
of
$\Sigma_{t_1}$ by $\eta=K\nu$. Since $G$ acts as isometries, we have $g_*\eta=
\eta\,\forall g\in G$ and hence $g\cdot exp_m(s\eta)=\exp_{g\cdot m}(s\eta)$
for sufficiently small parameters $s$ and $\forall m\in\Sigma_{t_1}$. Nearby
orbits are therefore labeled by $s$. Using $\eta$ as the deformation field
in the first variation formula of the area, we obtain:
$$
{d\over ds}\Bigl\vert_{s=0}A(s)\Bigr.=-\int_{\Sigma^*}K\langle \nu,
\eta\rangle\,\omega= -\int_{\Sigma^*}K^2\,\omega
$$
We conclude that stationarity with respect to the embedding parameter $s$
implies $K=0$}
imply stationarity amongst {\it all} variations and
hence that the image of the $t=\pi/6$ embedding of $S^3/D^*_8$ being a
minimal\note{
According to standard terminology, a stationary point of the area functional
is called a minimal surface and a stable minimal surface if it is a true
minimum.}
submanifold. We can see this directly by computing the extrinsic
curvature tensor $K=K_{ab}\theta^a\otimes\theta^b$ in the orthogonal frame
$$
\{\theta^a\}=\left\{dt\,,\,2\sin(t-\twothird\pi)\,l_1\,,\,
2\sin(t+\twothird\pi)\, l_2\, ,\, 2\sin (t)\,l_3\right\}
\eqno{(2.12)}
$$
The result is
$$\eqalignno{
\left\{K_{ab}(t)\right\}&=\hbox{diag}\left\{\cot (t-\twothird\pi),
\cot (t+\twothird\pi), \cot (t)\right\} &(2.13)\cr
tr\{K_{ab}(t)\}&=3\cot (t)\left[{\cot^2(t)-3\over 3\cot^2(t)-1}\right].
&(2.14) \cr}
$$

Note that the principal curvature directions (eigenvectors of $K_a^b$) are
just the left invariant vector fields, dual to the 1-form basis (2.12),
which, using (2.8), are readily seen\note{Standard formulae for the
Levi-Civita connection imply:
$(\nabla_XX,Y)=X(X,Y)-\shalf Y\vert X\vert^2-(X,[X,Y])$. Taking $X$ to be left
invariant gives trivially zero for all left invariant $Y$ and zero for
$Y={\partial\over\partial t}$ where the first term is zero and the second and
third just cancel. Hence $\nabla_XX=0$.} to be geodesic on $S^4$. Their
integral curves are thus great circles of $S^4$.

(2.13) implies $\{K_{ab}({\pi\over 6})\}= \hbox{diag}(0,\sqrt 3,-\sqrt 3)$
so that $tr\{K_{ab}\}({\pi\over 6})=0$ which means that $\Sigma_{\pi\over 6}$
is a minimal, non totally geodesic submanifold of $S^4$.

Although there is no isometry of
$S^4$ inducing the identity on $\Sigma_{\pi\over 6}$, there is
an isometry exchanging the two halves to either side of
$\Sigma_{\pi\over 6}$ orbitwise while inducing a non-trivial isometry
on $\Sigma_{\pi\over 6}$.
It is given by the standard antipodal map ${\cal A}$ of
$S^4$. To see this, first consider the map
$$
\eqalignno{
\sigma:\,[0,\pi/3]&\longrightarrow [0,\pi/3]\cr
                 t&\longmapsto {\pi\over 3}-t  &(2.15)\cr}
$$
whose action on the functions $\lambda_i$ in (1.12) is just given by.
$$\eqalignno{
\lambda_1\circ\sigma&=-\lambda_1\cr
\lambda_2\circ\sigma&=-\lambda_3&(2.16)\cr
\lambda_3\circ\sigma&=-\lambda_2\cr}
$$
On the other hand, consider the
following right multiplication map\note{Note that since
$\exp(-T_1\pi/2)\in N_{D^*_8}(SU(2))$ this is a
well defined map on the quotient $\Sigma$.}
$$\eqalignno{
R_{\exp (-T_1\pi/2)}:\,\Sigma &\longrightarrow\Sigma\cr
                      [p]&\longmapsto [p\cdot \exp(-T_1\pi/2)] &(2.17)\cr}
$$
where for $p\in S^3$, $[p]\in \Sigma$ denotes the equivalence class w.r.t.
right
multiplications by $D^*_8$. As is easily seen from (2.1) it acts on $A$ by just
exchanging the second and third eigenvalue of $\Lambda$. The composition of
(2.15) with (2.17) then acts on $A$ by multiplication with -1, which is just
the antipodal map. It clearly forms an isometry of (2.8).
This means that $\Sigma_{\pi\over 6}$ splits $S^4$ into two isometric halves.
Looking at eq. (2.8) we see that for $t\in\,]0,\pi/3[$ the only point
where two of the prefactors of the $l_i$'s coincide is $t=\pi/6$. This
implies that on a general $\Sigma_t$, the only isometries are the left
$SO(3)$ actions, whereas for $t=\pi/6$ there is an additional $Z_2$ factor
generated by the restriction of ${\cal A}$. This should be compared with
eq. (1.18).  Only a $Z_2$ subgroup of $P_3$ survived. Explicitly the metric
on $\Sigma_{\pi\over 6}$ is given by
$$
ds^2=4\,l_1^2+l_2^2+l_3^2
\eqno{(2.18)}
$$
On $S^3$ it has a $SU(2)$-left $U(1)$-right isometry group. On the
quotient space, $\Sigma$, this reduces to the $SO(3)$-left $Z_2$-right
isometries.

In the appendix it is shown that a general left invariant metric
$ds^2=\l_1^2 l_1^2+\l_2^2 l_2^2+\l_3^2 l_3^2$ has vanishing
Ricci-scalar if and only if $\l_i=\l_j+\l_k$ for some cyclic permutation
$i,j,k$ of $1.2.3$. But from $(2.7)$ it is immediate that this is
the case for all $\Sigma_t$. Note that since $(2.8)$ is the metric
on the unit four sphere, it satisfies
$$
R^{(4)}_{\mu\nu}=\Lambda g_{\mu\nu}\quad\hbox{where}\quad\Lambda=3\,.
\eqno{(2.19)}
$$
Performing a $(3+1)$ - decomposition for the purely normal components
of the Ricci-tensor with respect to the hypersurfaces $\Sigma_t$,
one obtains the corresponding Gauss-Codazzi equations (Hamiltonian
constraint)
$$
tr(K^2(t))+(tr(K(t)))^2=2\Lambda-R^{(3)}(t)=6
\quad\forall t\in]0,{\pi\over 3}[\,.
\eqno{(2.20)}
$$
where the last identity can also be explicitly checked from
$(2.13)$ using standard trigonometric identities and
$\Lambda=3$. In general relativity the left hand side of $(2.20)$
is just the so called kinetic term, which, except for its non
positive definite nature, formally represents the kinetic energy
per unit volume in the Hamiltonian. The `motion' described by
$t\mapsto\Sigma_t$ thus has `kinetic' energy proportional to
the volume $(2.9)$.

Let us finally have a look at Einstein's action principle.
The field equations $(2.19)$ can be derived from the action
functional
$$
S=\int_M (R^{(4)}-2\Lambda)\,\mu_M
  + 2\int_{\partial M} K_a^a\,\mu_{\partial M}
\eqno{(2.21)}
$$
where $R^{(4)}$ is the 4-dimensional Ricci-scalar and $\mu$
the standard measure induced by the metric. The boundary term is
necessary since the Ricci-scalar contains second derivatives
[3]. Let us now calculate
the action for a solution if $M$ is just half the 4-sphere,
cut along a minimal hypersurface $\Sigma=\partial M$. The action
for a solution is then just given by the integral of $\shalf R^{(4)}$
over $M$, which leads to $S= 8\pi^2 r^2$, where $r$ is the radius of
the 4-sphere. The boundary term in $(2.21)$ does not contribute due
to $\Sigma$ being minimal. In particular, if $V$ denotes the volume
of $\Sigma$, the action is proportional to $V^{2\over 3}$.
We can use this to compare the actions for different $\Sigma$ of equal
3-volume. Specifically, we compare the following two cases:
1.)~$\Sigma_{(1)}$ is the equatorial 3-sphere with volume
$V_{(1)}=2\pi^2r_{(1)}^3$,
2.)~$\Sigma_{(2)}=\Sigma_{\pi\over 6}$, as constructed above, with
volume $V_{(2)}=4\pi^2r_{(2)}^3$ (compare $(2.11)$). Since we want
$V_{(1)}=V_{(2)}$ we must have $r_{(2)}=2^{-{1\over 3}}r_{(1)}$
and hence for the actions:
$$
S_{(2)}=2^{-{2\over 3}}S_{(1)}=2^{-{4\over 3}}\,8(\pi V)^{2\over 3}
\eqno{(2.22)}
$$
For fixed 3-volume, the ``creation'' of the manifold $S^3/D_8^*$
costs only $2^{-{2\over 3}}$ times as much action as the
``creation'' of a round 3-sphere.

\beginsection{Appendix}

In this appendix we collect some formulae concerning the geometry
of the general left invariant metric
$$
ds^2= \l_1^2 l_1^2 + \l_2^2 l_2^2 + \l_3^2 l_3^2
    = \omega_1^2 + \omega_2^2 + \omega_3^2
\eqno{(A.1)}
$$
where $\{\l_1,\l_2,\l_3\}$ is the standard left invariant basis
as in $(2.5)$ which obeys $(2.6)$, and $\omega_i=\l_i l_i$ (no
summation). The normalization is such that the integral over
the unit 3-sphere of $l_1\wedge l_2\wedge l_3$ is $16\pi^2$.
This is easily verified using $(2.5)$ (note that
$0\leq\psi\leq 4\pi$). Solving Cartan's first structure equation
$d\omega_a+\omega_{ab}\wedge\omega_b=0$ for the antisymmetric
connection 1-forms $\omega_{ab}=-\omega_{ba}$ yields
$$
\omega_{12}=\eta_3\omega_3\quad\hbox{and cyclic}
\eqno{(A.2)}
$$
where `and cyclic' refers to any cyclic permutation of $1,2,3$,
and
$$
\eta_1={1\over 2}\left(-{\l_1\over\l_2\l_3}
       +{\l_2\over\l_3\l_1}+{\l_3\over\l_1\l_2}\right)
       \quad\hbox{and cyclic}\,.
\eqno{(A.3)}
$$
The curvature 2-form
$\Omega_{ab}=d\omega_{ab}+\omega_{ac}\wedge\omega_{cb}$
has then the non-vanishing components
$$
\Omega_{12}=(-\eta_1\eta_2 + \eta_2\eta_3 + \eta_3\eta_1)\,
  \omega_1\wedge\omega_2=\shalf R_{12ab}\,\omega_a\wedge\omega_b
\quad\hbox{and cyclic}
\eqno{(A.4)}
$$
The non-vanishing components of the Ricci-tensor are then the
diagonal ones:
$$\eqalign{
R_{11}
&= R_{2121}+R_{3131}=2\eta_2\eta_3 \quad\hbox{and cyclic}\cr
&= {1\over 2(\l_1\l_2\l_3)^2}\left(\l_1^4-(\l_2^2-\l_3^2)^2\right)
\quad\hbox{and cyclic}\cr}
\eqno{(A.5)}
$$
which leads to the Ricci-scalar
$$\eqalignno{
R= & R_{11}+R_{22}+R_{33}=2(\eta_1\eta_2+\eta_2\eta_3+\eta_3\eta_1)&\cr
 = & {1\over (\l_1\l_2\l_3)^2}\left(-(\l_1^2-\l_2^2-\l_3^2)
     + 4 \l_2^2\l_3^2\right)\quad\hbox{or cyclic}\,.&(A.6)\cr}
$$
In 3 dimensions the Ricci-tensor and Ricci-scalar determine the curvature.
According to the last two expressions it takes the simple
form:
$$
\Omega_{12}=(\shalf R-R_{33})\,\omega_1\wedge\omega_2
\quad\hbox{and cyclic}
\eqno{(A.7)}
$$

\proclaim
Observation. Since no two of the $\eta_i$ can simultaneously vanish, at
most one of the diagonal Ricci-components $R_{11},R_{22},R_{33}$ can
be zero. E.g. $R_{11}=0\Leftrightarrow \l_2^2=\l_1^2+\l_3^2$ or
(exclusively) $\l_3^2=\l_1^2+\l_2^2$. The Ricci-scalar vanishes, if and
only if $\l_a=\l_b+\l_c$ for some cyclic permutation
$a,b,c$ of $1,2,3$. If $R=0$, the Ricci-tensor is non-degenerate and
determines the sectional curvatures via $R_{1212}=-R_{33}$ and cyclic.

Let us briefly consider the case $R=0$. Without loss of generality we
assume $\l_1+\l_2=\l_3$. Then the non-vanishing components
for the Ricci-tensor and sectional curvature are:
$$\eqalignno{
& R_{11}=-R_{2323}={-2\over\l_2(\l_1+\l_2)} &(A.8a)\cr
& R_{22}=-R_{1313}={-2\over\l_1(\l_1+\l_2)} &(A.8b)\cr
& R_{33}=-R_{1212}={2\over\l_1\l_2}\,.      &(A.8c)\cr}
$$

\vskip1.0truecm
\centerline{\bf Acknowledgements}
I thank Gary Gibbons for initiating this study and
useful discussions.

\beginsection{References}

\item{[1]}
 General Relativity, An Einstein Centenary Survey. Edited by
 S.W. Hawking and W. Israel. Cambridge University Press, 1979.

\item{[2]}
M.A.H. MacCallum: Anisotropic and Inhomogeneous Relativistic
Cosmology. In [1].

\item{[3]}
S.W. Hawking: The Path Integral Approach to Quantum Gravity.
In [1].

\item{[4]}
S.W. Hawking (1984). The Quantum State of the Universe.
{\it Nucl. Phys. B} 239: 257-276.

\item{[5]}
G.W. Gibbons, J. Hartle (1991). Real Tunneling Geometries and the
Large-Scale Topology of the Universe.
{\it Phys. Rev. D}, {\bf 42}, 2458.

\bye